# Large anomalous Nernst effect in a van der Waals ferromagnet Fe$_3$GeTe$_2$

Jinsong Xu[1*], W. Adam Phelan[2], C.L. Chien[1]

[1] Department of Physics and Astronomy, Johns Hopkins University, Baltimore, Maryland 21218, USA

[2] Department of Chemistry, Johns Hopkins University, Baltimore, Maryland 21218, USA

**Abstract**

Anomalous Nernst effect, a result of charge current driven by temperature gradient, provides a probe of the topological nature of materials due to its sensitivity to the Berry curvature near the Fermi level. Fe$_3$GeTe$_2$, one important member of the recently discovered two-dimensional van der Waals magnetic materials, offers a unique platform for anomalous Nernst effect because of its metallic and topological nature. Here, we report the observation of large anomalous Nernst effect in Fe$_3$GeTe$_2$. The anomalous Hall angle and anomalous Nernst angle are about 0.07 and 0.09 respectively, far larger than those in common ferromagnets. By utilizing the Mott relation, these large angles indicate a large Berry curvature near the Fermi level, consistent with the recent proposal for Fe$_3$GeTe$_2$ as a topological nodal line semimetal candidate. Our work provides evidence of Fe$_3$GeTe$_2$ as a topological ferromagnet, and demonstrates the feasibility of using two-dimensional magnetic materials and their band topology for spin caloritronics applications.

*email: jxu94@jhu.edu

The recently discovered two-dimensional (2D) van der Waals (vdW) magnets [1-8], providing new constituent materials for spintronics and caloritronics, have attracted much attention. To date, most 2D vdW ferromagnets are semiconductors (e.g., $Cr_2Ge_2Te_6$) and insulators (e.g., $CrI_3$), with $Fe_3GeTe_2$ (FGT) being the only ferromagnetic (FM) metal. Electrical switching of FGT by spin orbit torque [9,10] and tunneling magnetoresistance in FGT/h-BN/FGT heterostruture [11] have recently been demonstrated. FGT also has the highest Curie temperature ($T_C$) among the 2D vdW ferromagnets, and the value of $T_C$ in thin FGT films can be tuned by ionic gating up to about room temperature [3]. Furthermore, the topological nature of FGT gives rise to some intriguing phenomena, for example, the observation of large anomalous Hall effect (AHE) [12] and magnetic skyrmions [13-15]. Of particular interest is the anomalous Nernst effect (ANE) in topological materials because the special band topology in these materials could introduce a very large ANE [16-19], which shares some similarities with, but also differences from, the better known AHE. In AHE and ANE, the current is driven by an electric field and temperature gradient $\nabla T$ respectively, while a voltage is measured perpendicular to both the current and the magnetization $M$ of the material. While the AHE is dominated by the sum of the Berry curvatures for all the occupied states, ANE is determined by the Berry curvature at the Fermi level $\epsilon_F$, thus providing a different probe of the Berry curvature near $\epsilon_F$ and the topological nature of materials [16-19]. Therefore, FGT provides a unique platform for studying the ANE in 2D vdW ferromagnets and their topological properties.

In this work, we report the observation of AHE and ANE in exfoliated FGT thin film devices. We have determined $\sigma_{xx}$ (defined as $\frac{\rho_{xx}}{\rho_{xx}^2+\rho_{xy}^2}$), and $\sigma_{xy}$ ($-\frac{\rho_{xy}}{\rho_{xx}^2+\rho_{xy}^2}$), the longitudinal and the transverse (or Hall) conductivities respectively, as well as $S_{xx}$ ($-\nabla V_x/\nabla T_x$) and $S_{yx}$ ($-\nabla V_y/\nabla T_x$), the longitudinal and the transverse Seebeck coefficients respectively. We have found very large AHE angle ($\theta_H = \sigma_{xy}/\sigma_{xx}$) and very large ANE angle ($\theta_N = S_{yx}/S_{xx}$), about 0.07 and 0.09 at low temperatures respectively, much larger than those found in common FM metals of 0.02 or less [12,20-23]. This places FGT outside the realm of common FM materials. We further found that the temperature dependence of the transverse thermoelectric conductivity $\alpha_{xy}$ follows the Mott relation $\alpha_{xy} = -(\frac{\pi^2 k_B^2}{3e})T(\frac{\partial \sigma_{xy}}{\partial \epsilon})_{\epsilon_F}$, and the Hall resistivity follows the scaling relation of $\rho_{xy} = \lambda M \rho_{xx}^n$ with $n \approx 2$, where $\lambda$ represents the strength of the spin-orbit coupling, $M$ the magnetization, and $\rho_{xx}$ ($\rho_{xy}$) the longitudinal (transverse) resistivity. This scaling relationship together with the observed large anomalous Hall conductivity (360~400 $\Omega^{-1}$ cm$^{-1}$) indicates the origin of the observed large AHE and ANE in FGT is primarily the intrinsic mechanism from the large Berry curvature, and not scattering, such as skew scattering and side-jump [23-25], consistent with the topological nature of FGT [12,26]. Our work provides evidence for FGT as a topological ferromagnetic metal, and demonstrates the feasibility of using 2D vdW magnetic materials and their band topology for spin caloritronic applications.

The vdW material of Fe$_3$GeTe$_2$ consists of Fe$_3$Ge slabs sandwiched by two layers of Te atoms and a van der Waals gap between the adjacent Te layers as shown in Fig. 1a. Single crystals of bulk FGT, grown by chemical vapor transport method, exhibit a lattice constant of $c$ = 16.36 Å, measured by X-ray diffraction (for details, see Supporting Information). The magnetic properties of bulk FGT, characterized by a superconducting quantum interference device, show a saturation magnetization of about 310 emu cm$^{-3}$ with a magnetic moment of 1.25 $\mu_B$ per Fe atom, and Curie temperature of about 200 K (see Supporting Information Fig. S3). Thin FGT flake is mechanically exfoliated onto heavily $n$-doped Si wafer with a 300 nm SiO$_2$ layer, then transferred onto pre-patterned Au contacts using dry transferred method (see Supporting Information for details). The thickness of our FGT samples are about 20~40 nm. Fig. 1b shows the schematic of a finished FGT device. There are two Au electrodes on the Si/SiO$_2$ substrate used as heaters to generate a lateral temperature gradient $\nabla T_x$ and also as a thermometer to measure the local temperature (see Supporting Information). There is a ~20 nm h-BN insulating layer to electrically isolate the Au heaters and the FGT layer. On top of h-BN, there are Au electrodes to make electrical contact to the FGT layer to measure the longitudinal voltage $V_x$ and transverse voltage $V_y$. For the measurements of both AHE and ANE, the magnetic field is applied out-of-plane, along the $z$-axis, unless otherwise noted.

We first examine the magnetic properties of FGT through the anomalous Hall effect with magnetometry results in the Supporting Information. For AHE, a small DC current (1 μA) is applied along the $x$-axis in FGT while both longitudinal voltage $V_x$ and transverse voltage $V_y$ are measured. We experimentally measured the longitudinal ($\rho_{xx}$) and the transverse ($\rho_{xy}$) resistivities, which convert to $\sigma_{xx} = \frac{\rho_{xx}}{\rho_{xx}^2 + \rho_{xy}^2}$ and $\sigma_{xy} = -\frac{\rho_{xy}}{\rho_{xx}^2 + \rho_{xy}^2}$, the longitudinal and the transverse (or Hall) conductivities respectively for quasi 2D materials. As shown in Fig. 2a, at low temperatures (50 K), the Hall conductivity $\sigma_{xy}$ shows a rectangular hysteresis loop with sharp switching under an external magnetic field $H_z$, suggesting a single magnetic domain over the entire device. Together with the near 100% remnant $\sigma_{xy}$ at zero field, these ferromagnetic hallmarks indicate thin FGT device exhibits strong perpendicular magnetic anisotropy (PMA) with the magnetic moments pointing in the out-of-plane direction. Thin films of common ferromagnets, hampered by the large shape anisotropy of *4πM*, rarely achieve PMA without very strong crystalline anisotropy or broken inversion symmetry and interface charge transfer. As the temperature increases (e.g., 140 K in Fig. 2a), the rectangular hysteresis loop gradually evolves into narrow-waist shape due to the weakened PMA and the formation of labyrinthine domain structure [4]. The temperature dependence of the Hall conductivity $\sigma_{xy}$ extrapolated to 0 Oe and the longitudinal conductivity $\sigma_{xx}$ is shown in Fig. 2b. For increasing temperature, the value of $\sigma_{xy}$ decreases, and vanishes around 200 K, similar to bulk single crystals of FGT. The longitudinal conductivity $\sigma_{xx}$ varies about 10% within the measured temperature range. Very significantly, the Hall angle $\theta_H = \sigma_{xy}/\sigma_{xx}$ reaches about 0.07 at low temperatures, similar to previous reports [12], much higher than those of common ferromagnets (≤0.02) [20-23].

We next describe the anomalous Nernst effect results in thin FGT devices. As depicted in Fig. 1b, by applying a DC current 14 mA to the right side heater, we generate an in-plane temperature gradient $\nabla T_x$ of about 1.3 K μm$^{-1}$. The transverse voltage $V_y$ is measured as a function of external perpendicular magnetic field $H_z$ as shown in Fig. 3a. Similar to that of AHE, we observe a rectangular hysteretic ANE loop. When we apply 12 mA current to the left side heater to generate an opposite temperature gradient $\nabla T_x$ of about -1.1 K μm$^{-1}$, we observe a similar but reversed rectangular hysteresis loop as shown in Fig. 3b. These results conclusively demonstrate ANE in the FGT device, after we address one experimental issue. It is well known that for thin films on thick substrates, the intended in-plane temperature gradient ($\nabla T_x$) is often accompanied inadvertently by an out-of-plane temperature gradient ($\nabla T_z$) due to the thermal conduction through the much thicker substrate. The latter may contribute an ANE voltage via $\nabla T_z \times M_x$, e.g., in permalloy thin films on Si substrate, where permalloy has in-plane magnetization $M_x$ [20]. To assess and eliminate this possible contribution, we apply a large magnetic field $H_x$ up to 6 kOe along the $x$ direction and observed no appreciable signal (see Supporting Information Fig. S5). This is because of the PMA of FGT is so strong that there is no measurable $M_x$ even with the presence of a large in-plane magnetic field. Thus all the measured thermal voltage is due to the anomalous Nernst signal of $\nabla T_x \times M_z$ in FGT.

To gain further insight and extract the Seebeck coefficients at zero field, provided by the robust PMA, we measure the ANE in the FGT devices at different temperatures $T^*$, and also with different temperature gradients $\nabla T_x$, as summarized in Fig. 3c and 3d. The longitudinal voltage $V_x$ and the transverse voltage $V_y$ are displayed in colors as a function of $T^*$ and $\nabla T_x$. Using the definitions of $S_{xx} = -\nabla V_x/\nabla T_x$ and $S_{yx} = -\nabla V_y/\nabla T_x$, we obtain the ANE angle $\theta_N = S_{yx}/S_{xx}$ via $LV_y/WV_x$, where $L = 10$ μm and $W = 5$ μm are the length and width of the sample respectively. We obtained very large ANE angle $\theta_N$ of as much as 0.09 at low temperature, a value much higher than those of common ferromagnets (<0.02) [20,21]. These large ANE angle $\theta_N$ of 0.09 corroborates with the similarly large AHE angle $\theta_H$ of 0.07, both are much larger than those in common FM metals of ≤ 0.02.

With the measured longitudinal ($S_{xx}$) and transverse ($S_{yx}$) Seebeck coefficients, we can obtain the transverse thermoelectric conductivity -$\alpha_{xy}$. The temperature dependence of the transverse and longitudinal Seebeck coefficient $S_{yx}$ and $S_{xx}$ is shown in Fig. 4a. The laws of thermodynamics and FM ordering dictate the temperature dependence of $S_{yx}$, which must vanish at $T = 0$ K, remains finite below $T_C$ and becomes negligible above $T_C$. With the aid of the Mott relation [16,23,27], quantitative analysis can be performed. The transverse thermoelectric conductivity -$\alpha_{xy}$ and Seebeck coefficients are related through

$$\alpha_{xy} = \sigma_{xx}S_{xy} + \sigma_{xy}S_{xx} = -\sigma_{xx}S_{yx} + \sigma_{xy}S_{xx}, \qquad (1)$$

Therefore, $\alpha_{xy}$ can be determined from the measured $\sigma_{xx}$, $\sigma_{xy}$, $S_{xx}$ and $S_{yx}$. On the other hand, if the Mott relation holds, $\alpha_{xy}$ can be obtained from via $\alpha_{xy} = -(\frac{\pi^2 k_B^2}{3e})T(\frac{\partial \sigma_{xy}}{\partial \epsilon})_{\epsilon_F}$, and $S_{xx}$ from $S_{xx} = -(\frac{\pi^2 k_B^2}{3e})T(\frac{\partial ln\sigma_{xx}}{\partial \epsilon})_{\epsilon_F}$, where $k_B$ is the Boltzmann constant, $\epsilon$ is the energy, and $\epsilon_F$ is the Fermi energy. By substituting the scaling relationship for the Hall resistivity $\rho_{xy} = \lambda M \rho_{xx}^n$ into the Mott relation and Eq. (1), we have [20,28]

$$S_{yx} = \sigma_{xy}/\sigma_{xx}\left(\left(\frac{\pi^2 k_B^2}{3e}\right)\frac{(\frac{\partial \lambda}{\partial \epsilon})_{\epsilon_F}}{\lambda}T + (n-1)S_{xx}\right), \qquad (2)$$

$$\alpha_{xy} = -\sigma_{xy}\left(\left(\frac{\pi^2 k_B^2}{3e}\right)\frac{(\frac{\partial \lambda}{\partial \epsilon})_{\epsilon_F}}{\lambda}T + (n-2)S_{xx}\right). \qquad (3)$$

To examine the Mott relation and determine the exponent $n$, we use Eq. (2) and Eq. (3) to fit $S_{yx}$ and $\alpha_{xy}$ to search for the best-fit values of $(\frac{\partial \lambda}{\partial \epsilon})_{\epsilon_F}/\lambda$ and $n$, as shown in Fig. 4. The best-fit exponent is $n = 2.3 \pm 0.1$ (blue solid curves). The exponent is clearly close to $n = 2$, and not $n = 1$, as shown in Fig. 4. The scaling relationship with $n = 2$ indicates that the intrinsic mechanism or side-jump, instead of skew scattering, dominates AHE and ANE [23-25] in FGT. Previous reports have shown the intrinsic mechanism dominates AHE in FGT [12,26], therefore *n* close to 2 favors intrinsic mechanism in FGT. At the same time, the anomalous Hall conductivity is $\sigma_{xy} \approx 360 \sim 400$ $\Omega^{-1}$ cm$^{-1}$ at low temperature, which is close to the intrinsic contribution $\sigma_{xy,in} \approx e^2/ha_z \approx 470$ $\Omega^{-1}$ cm$^{-1}$, where $h$ is Planck constant and $a_z$ is the interlayer distance ($a_z = c/2 = 8.18$ Å). This further corroborates the intrinsic Berry curvature as the primary source of the observed large AHE angle and ANE angle. This is consistent with the recent proposal of FGT as a topological nodal line semimetal with a large Berry curvature from the nodal line [12]. This may be the reasons for the exceptionally large AHE angle $\theta_H$ and ANE angle $\theta_N$ of 0.07 and 0.09, far larger than those in common ferromagnets.

In summary, we have observed ANE in exfoliated 2D van der Waals ferromagnetic metallic Fe$_3$GeTe$_2$ thin film devices. We have observed exceptionally large anomalous Nernst effect angle $\theta_N$ of 0.09, corroborated by the similarly large anomalous Hall effect angle $\theta_H$ of 0.07, much larger than the values of $\leq 0.02$ for common ferromagnetic metals. From the measured transport coefficients and the Mott relation, we find the large Berry curvature is responsible for the observed large AHE angle and ANE angle, which provides evidence for FGT as a topological ferromagnet. Our results demonstrate the feasibility of using 2D vdW magnetic materials and their band topology for spin caloritronics applications.

## ASSOCIATED CONTENT

**Supporting Information**

A description of device fabrication, $Fe_3GeTe_2$ bulk crystal growth and characterization, temperature calibration, and anomalous Nernst contribution from vertical temperature gradient.

## AUTHOR INFORMATION

**Corresponding Author**

*E-mail: jxu94@jhu.edu

**Authors contributions**

J.X. and C.L.C. conceived the experiment. W.A.P. synthesized the bulk crystals. J.X. performed the experiment. J.X. and C.L.C. wrote the manuscript with the input from all authors. All the authors discussed the results.

**Notes**

The authors declare no competing interests.

## ACKNOWLEGDEMENTS

This work was supported by the National Science Foundation under grant DMREF-1729555. The bulk crystals were grown in the Platform for the Accelerated Realization, Analysis, and Discovery of Interface Materials (PARADIM) facilities supported by the National Science Foundation under Cooperative Agreement No. DMR-1539918.

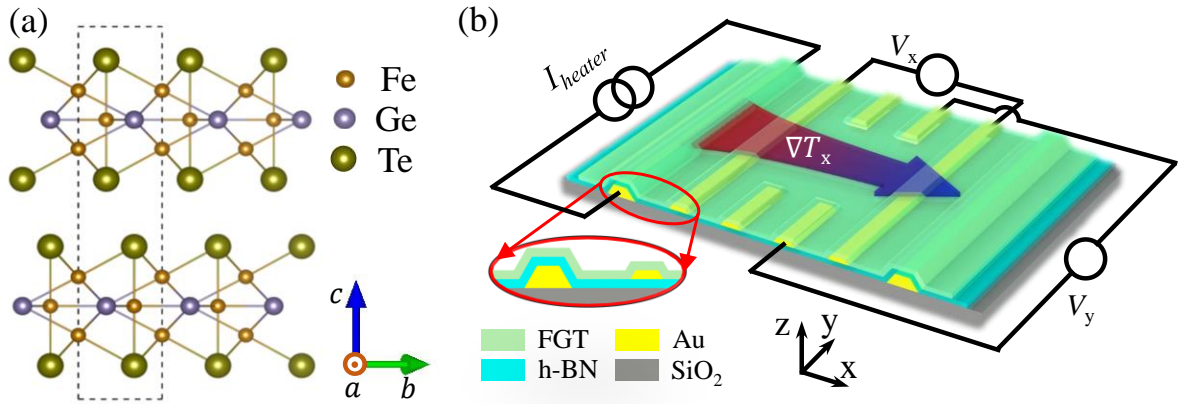

Figure 1. Crystal structure of Fe$_3$GeTe$_2$ (FGT) and FGT device for transport measurements. (a) Sideview of Fe$_3$GeTe$_2$ crystal structure. Fe$_3$Ge slabs sandwiched by two layers of Te atoms and a van der Waals gap between the adjacent Te layers. The dashed rectangular box denotes the unit cell. (b) Fe$_3$GeTe$_2$ device structure for anomalous Nernst effect. A lateral temperature gradient $\nabla T_x$ is applied along $x$ direction, a magnetic field $H_z$ is applied along $z$ direction, both longitudinal and transverse voltages $V_x$ and $V_y$ are measured.

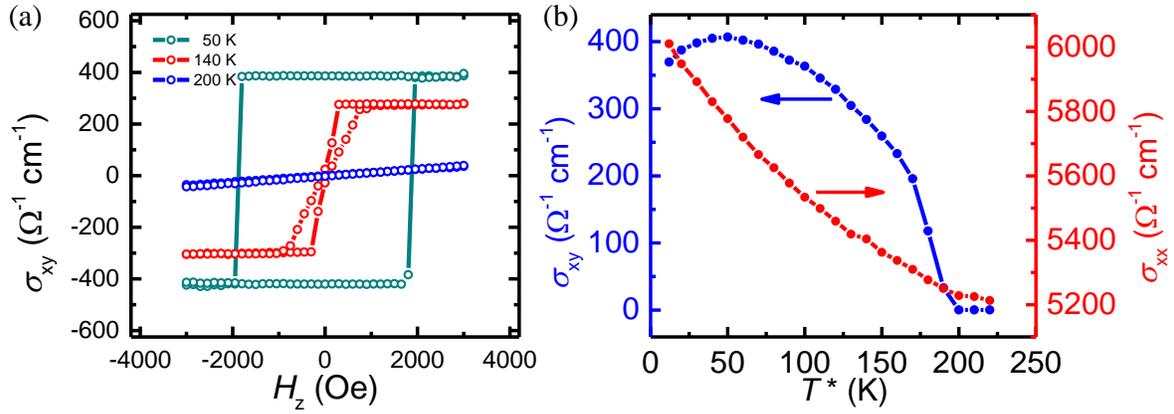

Figure 2. Temperature dependence of anomalous Hall effect (AHE). (a) Hall conductivity $\sigma_{xy}$ as a function of external applied magnetic field $H_z$ at different temperatures. (b) Temperature dependence of Hall conductivity $\sigma_{xy}$ (blue curve) extrapolated at zero field and longitudinal conductivity $\sigma_{xx}$ (red curve). $\sigma_{xy}$ vanishes around 200 K, the Curie temperature $T_C$ of the sample. The Hall angle $\theta_H = \sigma_{xy}/\sigma_{xx}$ reaches about 0.07 at low temperature.

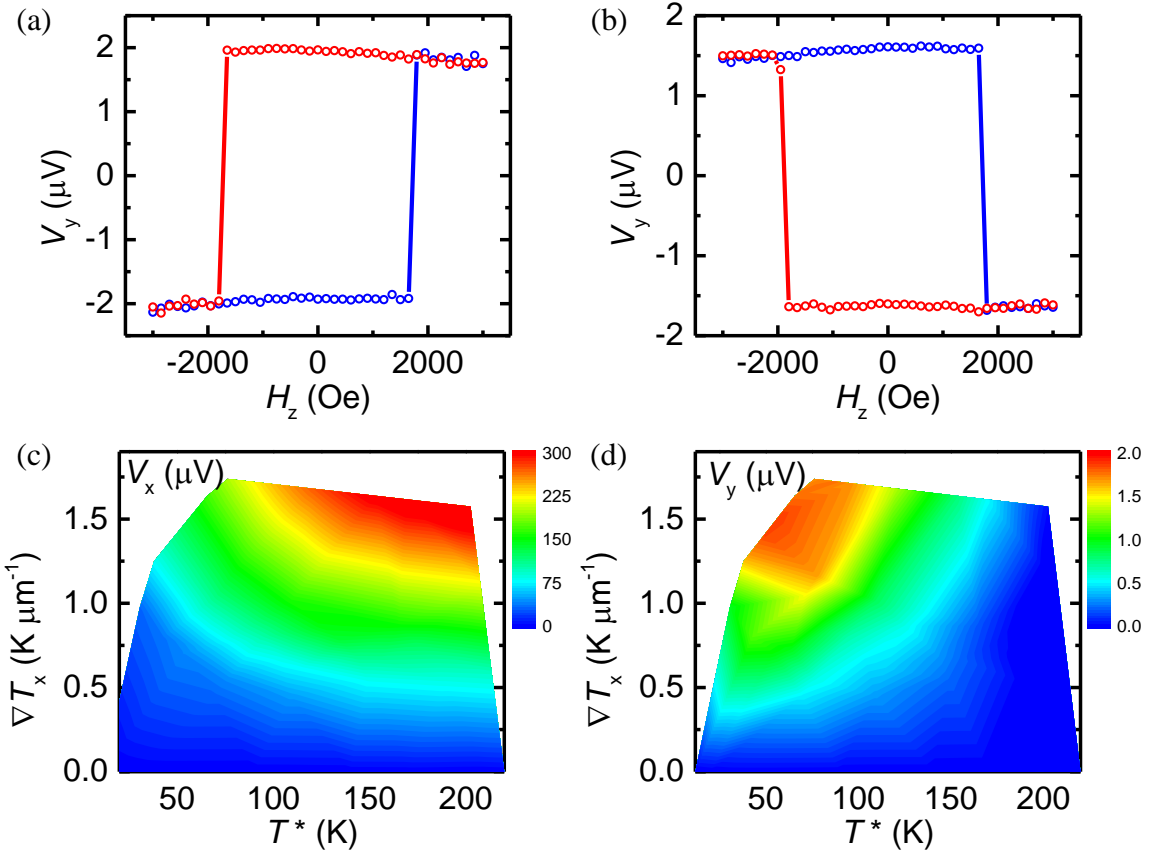

Figure 3. Anomalous Nernst effect (ANE), longitudinal voltage ($V_x$) and transverse voltage ($V_y$). (a) Nernst signal $V_y$ as a function of $H_z$ at $\nabla T_x$ =1.3 K μm$^{-1}$. (b) Nernst signal $V_y$ as a function of $H_z$ at $\nabla T_x$ =-1.1 K μm$^{-1}$. The data are taken at an effective sample temperature $T^* = 45$ K. The blue (red) curve is for increasing (decreasing) magnetic field. (c) Longitudinal voltage $V_x$ (displayed in colors) as a function of temperature and temperature gradient. (d) Transverse voltage $V_y$ (displayed in colors) as a function of temperature and temperature gradient. By definition, $\nabla V_x = - S_{xx}\nabla T_x$ and $\nabla V_y = - S_{yx}\nabla T_x$, both $V_x$ and $V_y$ will increase as a function of $\nabla T_x$. The general trend that $V_x$ ($V_y$) increases (then decreases) as $T^*$ increases, indicate $S_{xx}$ ($S_{yx}$) increases (then decreases) with sample temperature $T^*$.

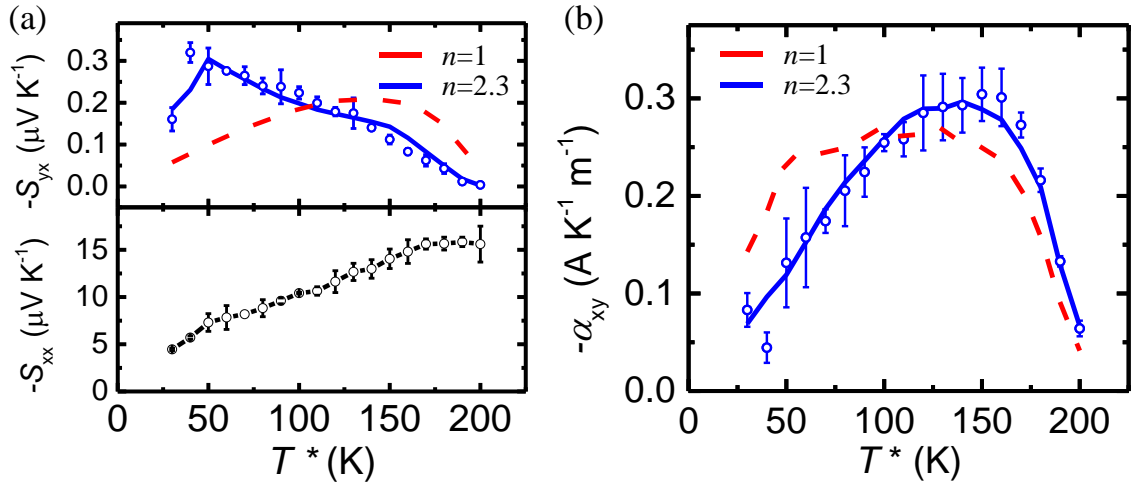

Figure 4. Temperature dependence of $S_{yx}$, $S_{xx}$ and $\alpha_{xy}$. (a) Temperature dependence of $S_{yx}$ (top panel) and $S_{xx}$ (bottom panel). $S_{xx}$ and $S_{yx}$ is extracted from the linear fitting $V_x$ and $V_y$ vs $\nabla T_x$ respectively. (b) Temperature dependence of $\alpha_{xy}$. $\alpha_{xy}$ is determined via Eq. (1). Blue solid curve is the best fit using Eq. (2) and (3) for $S_{yx}$ and $\boldsymbol{\alpha_{xy}}$ respectively, and red dashed curve is the best fit with $n=1$.

# Supporting Information for

# Large anomalous Nernst effect in a van der Waals ferromagnet Fe$_3$GeTe$_2$

Jinsong Xu[1], W. Adam Phelan[2], C.L. Chien[1]

[1] Department of Physics and Astronomy, Johns Hopkins University, Baltimore, Maryland 21218, USA

[2] Department of Chemistry, Johns Hopkins University, Baltimore, Maryland 21218, USA

## 1. Device fabrication

A dry transfer technique was used to fabricate thin FGT film device for the ANE study. Thin flakes of FGT and h-BN were mechanically exfoliated from their bulk crystals onto Si substrates covered by a 300 nm thermal oxide layer. First, we used standard e-beam lithography with MMA/PMMA bilayer resist to fabricate local heater electrodes. Au electrodes (40 nm) were deposited on Si/SiO$_2$ (300 nm) empty wafer using an e-beam source and a 5 nm Ti underlayer for adhesion (Fig.S1 (a)). Then a stamp, which consisted of a thin layer of poly(propylene carbonate) (PPC) on polydimethylsiloxane (PDMS) supported by a glass slide, was used to pick up an exfoliated h-BN flake (~20 nm) and transfer onto pre-patterned local heater (Fig.S1 (b)). After that, a second e-beam lithography was used to fabricate contact electrodes. Au electrodes (20 nm) were deposited on Si/SiO$_2$ (300 nm)/h-BN (20 nm) with a 3 nm Ti underlayer for adhesion (Fig.S1 (c)). Finally, another PPC stamp was used to pick up an exfoliated FGT flake (20~40 nm) and transfer onto the pre-patterned contact electrodes (Fig.S1 (d)).

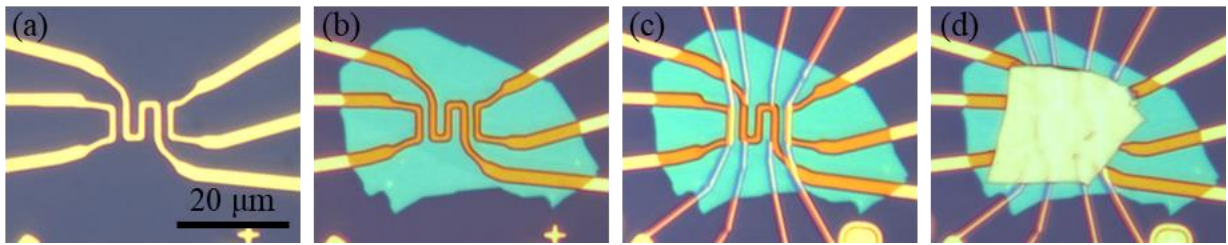

Figure S1. Optical images of FGT devices. (a) Local heaters on Si/SiO$_2$ substrate. (b) h-BN on top of local heaters. (c) Au electrodes for electrical contact on h-BN. (d) FGT transferred onto prepatterned electrodes. The scale bar is 20 μm.

## 2. FGT bulk crystal growth

High-quality FGT single crystals were grown by the chemical vapor transport method. High-purity elements were stoichiometrically mixed and sealed under vacuum in a quartz tube with a small quantity of iodine. The tube was placed into a two-zone horizontal furnace with the hot end at 750 °C and the cold end at 650 °C. Shiny plate-like single crystals with typical dimensions of 3 mm × 3 mm × 0.25 mm were then obtained. The high quality of the bulk materials is confirmed by X-ray diffraction (XRD) (Fig.S2) and superconducting quantum interference device (SQUID) (Fig.S3). The c-axis lattice constant extracted from the XRD data in Fig.S2 (a) is 16.36 Å. And the XRD φ scan along (016) plane shows six-fold symmetry (Fig.S2 (b)). The magnetization measurement show that the saturation magnetization is about 310 emu cm$^{-3}$ with a magnetic moment of 1.25 $\mu_B$ per Fe atom (Fig.S3 (a)), and Curie temperature is about 200 K (Fig.S3 (b)).

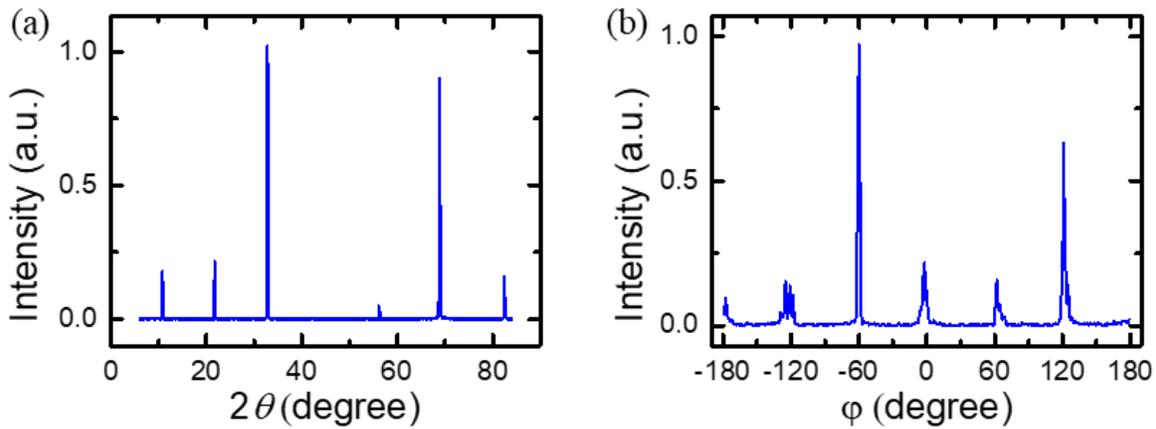

Figure S2. XRD on single crystal FGT. (a) XRD 2θ-ω scan of FGT (001) planes. (b) XRD φ scan of FGT (016) plane.

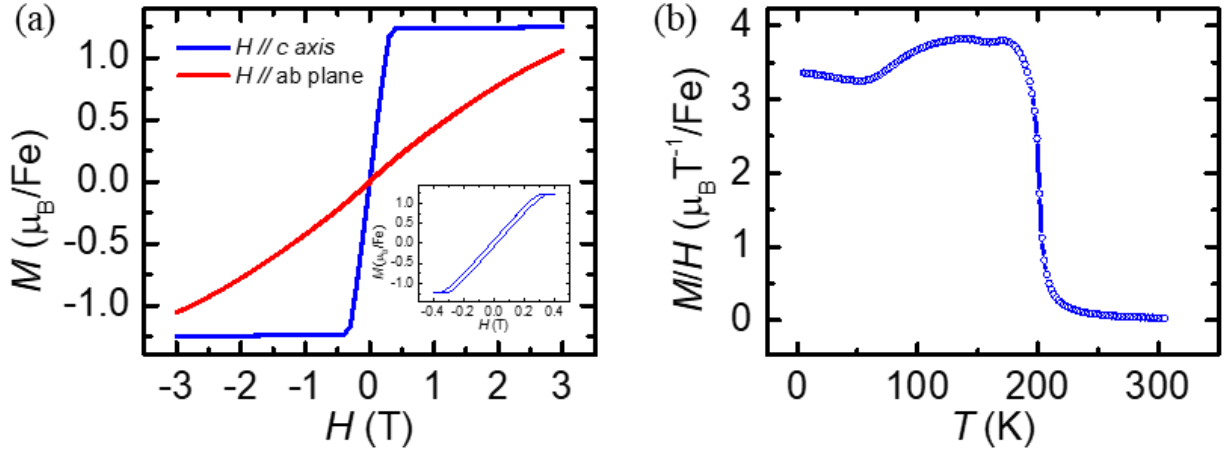

Figure S3. Magnetic properties of single crystal FGT. (a) Magnetization loops of bulk FGT at 5 K. Inset is the zoom-in around 0 field for $H$ applied along c axis. (b) Temperature dependence of the magnetization $M$ divided by applied field $H$, $H$ is fixed at 100 Oe along c-axis.

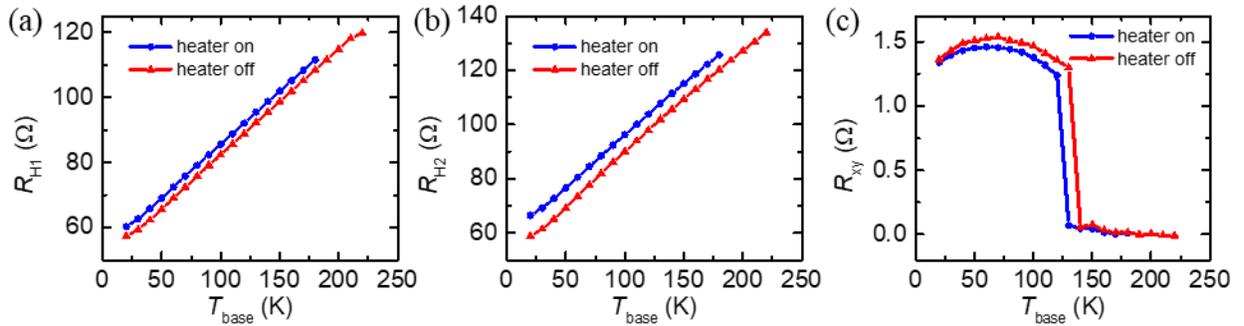

Figure S4. Temperature dependence of heater resistance and AHE. (a) Temperature dependence of heater H1 resistance. (b) Temperature dependence of heater H2 resistance. (c) Temperature dependence of AHE. Blue curves are resistance when heater is off, and red curves are resistance when 10 mA current is applied to heater H2.

## 3. Temperature calibration

The local temperature of the heater is calibrated by the heater resistance, and the effective sample temperature $T^*$ is calibrated by the AHE signal. $T_{base}$ is the temperature of the cryostat. First, we apply a small current (1 μA) to measure the heater resistance and AHE. In this case, the heating effect is negligible, the temperature of the heater and sample is basically the same to $T_{base}$. These three temperature dependence of resistance curves (blue curves in Fig.S4) are used as calibration curves to measure the local temperatures. For example, when 10 mA current is applied to heater H2 (red curves in Fig.S4), local temperature of heater H1 $T_{H1} \approx T_{base} + 12$ K, $T_{H2} \approx T_{base} + 20$ K, and $T^* \approx T_{base} + 14$ K.

## 4. Nernst signal contribution from vertical temperature gradient $\nabla T_z$

Because the anomalous Nernst signal $V$ is proportional to $\nabla T \times M$, a voltage signal along $y$-axis $V_y$ could be generated by $\nabla T_x \times M_z$ or $\nabla T_z \times M_x$. To eliminate the possible contribution from the vertical temperature gradient $\nabla T_z$, we measure the transverse voltage $V_y$ as a function of external applied magnetic field $H_x$ while keeping the same heater current (14 mA) as used in the main text in Fig. 3a. As seen in Fig. S5, there is no appreciable Nernst signal with fields less than 3000 Oe. Therefore, the Nernst signal in the main text is from lateral temperature gradient $\nabla T_x$.

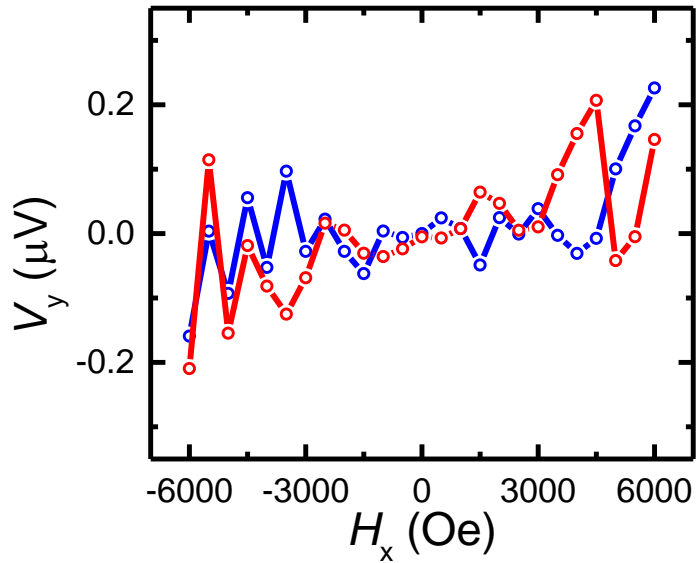

Figure S5. Transverse voltage $V_y$ as a function of external applied magnetic field $H_x$. The blue (red) curve is for increasing (decreasing) magnetic field.